\documentstyle[NATO,epsfig,graphics]{crckapb}
\vspace{4cm}
\begin{opening}
\title { Local Charge distributions in Metallic Alloys: a Local Field Coherent
Potential Approximation Theory}
\author{Ezio Bruno}
\author{Leon Zingales} 
\author{Antonio Milici} 
\institute{\\ Dipartimento di Fisica and Unit{\`{a}} INFM,
Universit{\`{a}} di Messina,  Salita Sperone 31, 98166 Messina, Italy.
 E-mail: bruno@dsme01.unime.it. }
\end{opening}
\begin{document}
\section{INTRODUCTION}

In the last decade order N electronic structure calculations~\cite{LSMS,FWS2} made possible the study of large
supercells containing from 100 to 1000 atoms. Namely Faulkner, Wang and
Stocks~\cite{FWS2,FWS1} have shown that simple linear laws, the so called
'$qV$'  relations, link the local charge excesses and the local Madelung
potentials in metallic alloys. These $qV$ linear laws have been obtained
from the numerical analysis of data produced by Locally 
Self-consistent Multiple Scattering (LSMS)~\cite{LSMS} calculations, while their formal derivation within the density functional theory
has not yet been obtained. As a matter of fact, the above laws can be considered to hold at least
within the approximations underlying LSMS calculations, i.e.
the Local Density and the muffin-tin
approximations. 

In this paper we shall develop a new version of Coherent Potential Approximation
theory (CPA). We apply a local
external field and study the response of the mean field CPA alloy.
Because of the fluctuation-dissipation theorem, the response to the
external field must be equal to the internal field caused by
electrostatic interactions. This new theoretical scheme, avoiding the
consideration of specific supercells, will enable us to explore a broad range of
fields and clarify certain aspects of the mentioned $qV$ relations.

We shall find that, in a quite broad range of applied fields, $\Phi$, the 
{\it integrated} charge excess at a given site, $q$, scales linearly with the 
field, in agreement with the findings of Refs~\cite{FWS2,FWS1}. However, 
remarkably, in the same range of $\Phi$ values, the charge density at a 
given point does not obey a linear scaling. Our results
for the CuPd and CuZn alloy systems compare favourably both with the LSMS and
conventional superlattice multiple scattering theory calculations, as well
as with the available experimental data.
Our theory, when applied to random alloys, is computationally inexpensive in
comparison with other approaches and can, in principle, be used, in
conjunction with statistical methods, to describe ordering phenomena in
metallic alloys.

In the next section 2, we shall discuss about charge transfers in multiple
scattering theory calculations and CPA theory, while in section 3 we
shall describe the above new version of the CPA theory that incorporates
local fields (CPA+LF) and apply
it to the study of fcc CuPd and fcc and bcc CuZn alloys. In the
conclusion we shall summarize the most important features of our
work.
\newpage
\setlength{\topmargin}{-1.3cm}
\setlength{\textheight}{25.5cm}
\section{CHARGE TRANSFERS IN METALLIC ALLOYS.}

\subsection{Charge transfers from  LSMS calculations.}

Faulkner, Wang and Stocks~\cite{FWS2,FWS1} have
analysed the
distribution of charges in binary metallic alloys as obtained from LSMS 
calculations. They have studied large
supercells with periodic boundary conditions containing
 hundreds of atoms and designed to simulate substitutional
disorder. LSMS calculations are based on the local density approximation to the density functional
theory~\cite{HK&KS,Dreizler} and the muffin-tin approximation for the crystal potential; thus the
 results of their analysis hold $\it{within}$ the same approximations. Below we
 shall summarize and comment the conclusions obtained in Refs.~\cite{FWS2,FWS1} that are relevant
for our present
concerns: 

i) { \it For a given alloy configuration}, the site charges $q_i$ and the Madelung potentials $V_i$ 
obtained from LSMS calculations for binary alloys {\it lie on two straight lines} of equations:
\begin{equation}
\label{qvsv}
a_i q_i + V_i = k_i
\end{equation}
where the quantities $a_i$ and $k_i$ take the values $a_A$ and $k_A$ if the i-th site is occupied by
an A atom and $a_B$ and $k_B$ if it is occupied by B. The size of the deviations from linearity appears
 $\it{comparable}$ with the numerical accuracy of LSMS calculations. 

The Madelung potentials $V_i$ entering in Eq.~(\ref{qvsv}) are determined by the charges at all the crystal
 sites through the relationship:
\begin{equation}
\label{Mad}
V_i = 2 \sum_i M_{ij} q_j
\end{equation}
where the factor 2 comes from using atomic units. The Madelung matrix
elements, $M_{ij}$, are defined~\cite{Ziman} as
\begin{equation}
\label{madmat}
M_{ij}=\sum_{\vec{R}} \frac{1}{|
\vec{r}_{ij}+\vec{R}|}
\end{equation}
in terms of the translation vectors from the
i-th to the j-th site, $\vec{r}_{ij}$, and the supercell lattice vectors $\vec{R}$. 

ii) For different alloy configurations corresponding {\it to the same molar 
fractions}, the four
constants $a_A$, $k_A$, $a_B$ and $k_B$ in Eq.~(\ref{qvsv}) take different values. This notwithstanding, the
variations of the same constants when considering different samples at the same concentration
appear much smaller than their variation with the concentration. 

iii) The site charge excess corresponding to each chemical species in a random alloy configuration
take {\it any} possible value in some interval $q_{min} \leq q_i \leq q_{max}$. 

Faulkner, Wang and Stocks~\cite{FWS2,FWS1} have stressed that the existence of a linear relation is not a
 trivial consequence of classical electrostatics. In fact, Eq.~(\ref{qvsv}) is not verified at a
generic Kohn-Sham iteration for the charge density in LSMS calculations, while it is found to hold
{\it only} when convergence is achieved. Thus the linearity of the $qV$ laws should be interpreted
as a consequence of the screening phenomena that occur in metals. As shown by Pinski~\cite{pinski},
linear $qV$ laws can be obtained also by Thomas-Fermi density functional calculations. This
circumstance strongly suggests that the linearity of the $qV$ relations has little to do with the
specific form of the density functional used in the calculation. The conclusions drawn in Refs.~\cite{FWS2,FWS1}
and summarized above are indirectly supported by photoemission experiments~\cite{Weightman,FWS3}.
Moreover, electronic structure calculations based on the Locally Self-consistent Green's
Function method (LSGF) and the atomic sphere approximation for the crystal potential have also
confirmed the linearity of the $qV$ relations~\cite{Abrikosov_cpa,Ruban1,Ruban2}. 

It should be clear
that the definition of charge excess is based on the quite artificial 
partition of the crystal volume
into "atomic" sites. This partition is accomplished using the muffin-tin approximation
in Refs.~\cite{FWS2,FWS1} or the atomic sphere
approximation in Refs.~\cite{Abrikosov_cpa,Ruban1,Ruban2}. Of course other procedures are possible,
but even
in the case in which no spherical approximation is made, as it could be for full potential
calculations (that unfortunately are not yet available), the way in which the "atomic cells" are chosen
would remain arbitrary. However, different partitions of the crystal volume {\it always} lead to linear
laws. This has been shown, e.g., in Ref.~\cite{Ruban1} by changing the ratio $r$ between the atomic radii associated
with each chemical species~\cite{note}. To summarize: at least when a spherical
approximation is used, the functional form of Eq.~(\ref{qvsv}) is maintained while, of course,
the actual values
of the coefficients depend on the particular partition used.

As it is evident, the presence of the charge transfers leads to energy corrections that can
be important in the physics of metallic alloys. The
simple functional form in Eq.~(\ref{qvsv}) allows an easy route for including such corrections~\cite{BMZc}.
An alternative way for accounting the electrostatic energy contribution due to charge transfers has been
proposed by Gonis et al~\cite{note,gonissb}.
It consists in choosing the dimensions of the atomic spheres for each alloying
species in such a way to have zero charge transfers and, hence, zero contribution to the total energy. 
Of course, such a procedure could cause large overlap volumes
(for simple lattices the overlap volume is minimum when equal atomic spheres are used) and,
hence, large errors in density functional theory calculations. 

Although, in principle, the quantities $a_i$, $k_i$ in Eq.~(\ref{qvsv}) can be influenced by the local
environments, it is clear that {\it the consideration of the site chemical occupation only is
sufficient to determine the same quantities} within an accuracy comparable with the numerical errors
in LSMS calculations. This circumstance, as a matter of fact, suggests that a single site theory~\cite{VKE}
as the CPA could be sufficient to determine the above $a_i$, $k_i$. In 
section 3 this suggestion
shall be analysed.

\subsection{Charge transfers in the CPA theory.}
For many years the CPA theory~\cite{Soven} has been used for calculating the
electronic properties of random metallic alloys.
In fact, the CPA has allowed for very careful studies of spectral
 properties~\cite{Abrikosov_cpa}, Fermi surfaces~\cite{physrep}, phase equilibria~\cite{phase} and magnetic
 phenomena ~\cite{magnetic} in metallic alloys. Moreover, in spite of its 
 simplicity, the theory has achieved remarkable
successes in the calculation of properties related with Fermi liquid effects,
such as spin~\cite{spinwaves}
 and concentration waves~\cite{cwaves}. However, for the purpose of the present work two aspects of the theory are particularly relevant:
its elegant formulation
in terms of multiple scattering theory~\cite{Gyorffy,KKR} and the fact that it constitutes the natural
 first step for perturbative studies. 

As it is well known~\cite{Magri}, the CPA does not include the energetic
contributions that derive from charge transfers in metallic alloys. In
spite of this, the CPA is useful for understanding some physical properties
related with these charge transfers. We will try to explain below the reasons 
for this apparent paradox.

The CPA theory (we shall use the multiple scattering theory
formalism~\cite{Gyorffy,KKRCPA} for a random binary alloy $A_{c_A}B_{c_B}$)
consists of solving for $t_C$ the so called CPA equation,
\begin{equation}
\label{CPA}
c_A G_A(t_A,t_C)+c_B G_B(t_B,t_C)=G_C(t_C)
\end{equation}

\begin{figure}
\centerline{\epsfig{figure=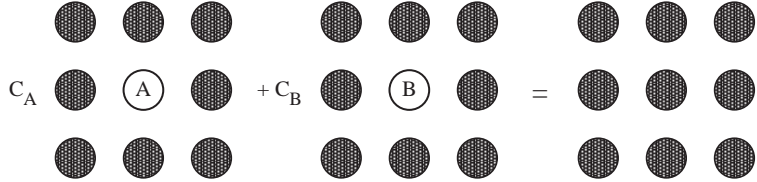,width=10cm}}
\vspace{0.5cm}
\caption{Schematic illustration of the CPA theory. Dark sites are
occupied by the CPA coherent scatterer described by the single site
scattering matrix $t_C$. The central impurity sites, labelled by A and B,
are characterised by the single site matrices $t_A$ and $t_B$.}
\label{cpa}
\end{figure}

The three Green's functions in Eq.~(\ref{CPA}), $G_A(t_A,t_C)$,
$G_B(t_B,t_C)$ and $G_C(t_C)$, refer to the three different
problems sketched in Fig.~\ref{cpa}. In fact, $G_C(t_C)$ is the Green's function
for an infinite crystal whose sites are all occupied by effective scatterers
characterised by the single-site scattering matrix $t_C$. 
On the other
hand, $G_{A(B)}(t_{A(B)},t_C)$ is the Green's function for a single impurity
'atom' described by the single-site scattering matrix $t_{A(B)}$ and embedded in
an infinite crystal with all the other sites
characterised by the single-site scattering matrix $t_C$.

While the homogeneous effective crystal, the 'coherent' medium of the CPA
theory, let us call it C, is electroneutral, the two impurity Green's
functions lead to net charge excesses, $q^0_A$ and $q^0_B$, in the sites occupied
by the A and B impurities. On behalf of Eq.~(\ref{CPA}), these charge excesses
satisfy the condition,
\begin{equation}
\label{elecn}
c_A q^0_A + c_B q^0_B=0
\end{equation}
In C there is no charge transfers from one site to the others and, thus,
Eq.(\ref{elecn})
cannot be interpreted as an ordinary electroneutrality condition.
However  $q^0_{A(B)}$ can be considered as the charge that the impurity
A(B) attracts from the mean medium C, in the sense that there is an
indirect charge transfer from A to B, through the mean medium C. The last can be
reinterpreted as a reference system and plays a role similar to that of
the Hydrogen atom for molecules, in the formulation of the electronegativity
theory by Pauling~\cite{Pauling}. 

In summary: we could say that the CPA 'charge transfers' $q^0_{A(B)}$ {\it reflect} the difference
of electronegativity between A and B.
Of course the CPA theory, being a single site and a mean field theory, cannot account
for the complex charge relaxation phenomena that are expected
to make non equivalent sites occupied by same species and surrounded by
different local chemical environments.
In order to have a picture in which sites occupied
by the atoms of the same kind are no longer equivalent, it is necessary
to  renounce to a single-site picture. Non single-site formulations of the
CPA theory have been proposed several times in the literature. Here we mention
the charge-correlated CPA by Johnson and Pinski~\cite{more_scr} and the Polymorphous Coherent
Potential Approximation (PCPA) by Ujfalussy et al~\cite{Ujfalussy}. In this paper, we shall
 develop a different approach by
introducing an external local field in a single-site CPA picture; this will allow to
maintain all the mathematical simplicity of a
single-site theory, nevertheless the presence of external fields will be sufficient to lead to
polymorphous site potentials.

\section{RESPONSE TO LOCAL FIELDS OF THE 'CPA ALLOY'.}
\subsection{The local field CPA (CPA+LF) model}

In this section, we develop a new version of CPA theory by introducing an
external local field  $\Phi$. It will formally enter in the theory
as a parameter that can be varied at will. We
shall focus on the response of the system due to the resulting rearrangement of the
charge distribution. 

\begin{figure}
\centerline{\epsfig{figure=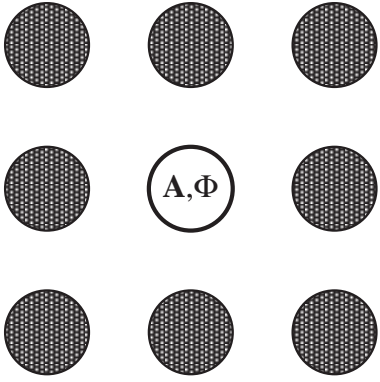,width=5cm}}
\vspace {0.5cm}
\caption{Schematic illustration of the CPA+LF method. As in Fig.~\ref{cpa},
dark sites are occupied by the CPA coherent scatterer described by $t_C$.
In the central site, occupied by A, acts also a constant field $\Phi$.}
\label{cpaphi}
\end{figure}

We imagine an A impurity atom in a otherwise
homogeneous crystal with all the other sites occupied by C scatterers. We
suppose that the single site scattering matrix of the CPA  medium, $t_C$,
and its Fermi energy, $E_F$, have been determined by the CPA theory for the
binary alloy $A_{c_A}B_{c_B}$. The local external field, $\Phi$,
takes a constant value within the impurity site volume and is zero
elsewhere~\cite{note2}. This situation is pictorially represented in
Fig.~\ref{cpaphi}. To simplify our discussion we shall solve the problem using
the Atomic Sphere Approximation (ASA). However, the following
considerations hold for any cellular method,
and, with minor modifications, also for the muffin-tin approximation.

We shall refer to the impurity A in the presence of the external field
$\Phi$ as to $(A,\Phi)$. When $\Phi=0$, the site Green's function associated
with it, $G^\Phi_A(t^\Phi_A,t_C)$, reduces to the usual CPA Green's function,
$G_A(t_A,t_C)$. When $\Phi\ne 0$, $G^\Phi_A(t^\Phi_A,t_C)$ can be readily
obtained using the multiple scattering theory impurity
formula ~\cite{KKRCPA}:

\begin{eqnarray}
\label{Green}
G^\Phi_A(E, \vec{r}, \vec{r}^{\;\prime})  & = &
\sum_{L,L^\prime} [ Z^\Phi_L(E, \vec{r})
\tau^\Phi_{A,LL^\prime} Z^\Phi_{L^\prime} (E, \vec{r}^{\;\prime}) -
  Z^\Phi_L(E,
\vec{r}) J^\Phi_{L^\prime} (E, \vec{r}^{\;\prime}) \delta_{LL^\prime}]
\end{eqnarray}
where
\begin{equation}
\label{tau}
\tau^\Phi_A=D^\Phi_A \tau_C=\left[1+\tau_C \left( (t^\Phi_A)^{-1}-t_C^{-1}
\right) \right]^{-1} \tau_C
\end{equation}
In Eqs.~(\ref{Green}) and (\ref{tau}), E is the energy, $t_C$ and $\tau_C$ are
the CPA single site scattering matrix and scattering-path operator,
{\it as determined by a standard CPA calculation}, i.e. $\Phi=0$,
for the alloy at hand. The single site scattering matrix corresponding
to $(A,\Phi)$, $t^\Phi_A$, is to be determined from the site
potential $V^\Phi_A(\vec{r})+\Phi$, $D^\Phi_A$ is the CPA projector relative
to the same site potential, $Z^\Phi_L(E, \vec{r})$ and $J^\Phi_L(E, \vec{r})$
are two orthogonal solutions of the Schroedinger equation for the same potential,
the first of which is regular at $r=0$. In our notation
$L=(l,m)$ labels the angular momentum quantum numbers and, for sake of
simplicity, we omit the energy dependence of all the scattering matrices.
A complete account of the notation can be found in Ref.~\cite{KKRCPA}.

The charge density corresponding to $(A,\Phi)$ is obtained integrating
Eq.~(\ref{Green}) over the energy up to the Fermi level,
\begin{equation}
\label{rhoel}
\rho^\Phi_A(\vec{r})=-\frac{1}{\pi} Im \; \bigg\{ \int_{-\infty}^{E_F} dE \;
 G^\Phi_A(E, \vec{r}, \vec{r}^{\;\prime}=\vec{r}) \bigg\}
\end{equation}

The corresponding site potential, $V^\Phi_A(\vec{r})$, can be reconstructed
by solving the appropriate Poisson equation and adding the exchange-correlation
contribution~\cite{Janak,Winter}. Unless $\Phi=0$, it will be
different from the site potential obtained from the zero field CPA theory,
$V_A(\vec{r})=V^{\Phi=0}_A(\vec{r})$, due to the charge relaxations
expected to screen in part the external field. In a numerical implementation of the theory, Eqs.~(\ref{Green}-\ref{rhoel}) and
the potential reconstruction need to be iterated starting from a convenient
initial guess, until convergence is achieved for
$V^\Phi_A(\vec{r})$ or, equivalently, for $\rho^\Phi_A(\vec{r})$. Hereafter we shall
refer to the above model as to the Local Field CPA (CPA+LF).

Once convergence is obtained for the charge density, the net charge on the
site A
can be obtained by integrating over the atomic sphere volume and subtracting
the nuclear charge, $Z_A$,
\begin{equation}
\label{charge}
q_A(\Phi)=\int d\vec{r} \rho^\Phi_A(\vec{r}) - Z_A
\end{equation}

It is important to realise that, while the above self-consistent
determination of $V^\Phi_A(\vec{r})$ or $\rho^\Phi_A(\vec{r})$ allows for
full charge relaxation at the impurity site, the CPA+LF does not modify the
properties of the CPA medium C: these remain specified by the quantities $t_C$
and $E_F$ determined at zero external field. The resulting lack of self-consistency in the CPA+LF
is not a serious drawback if one is interested, as in the present case, to the investigation of trends
and general aspects of the screening phenomena.

\subsection{CPA+LF results for CuZn and CuPd alloys: the site charges}
We have implemented the CPA+LF theory within our well tested KKR-CPA
code~\cite{KKRAlgorithms}. If $t_C$ and $\tau_C$ from a previous standard KKR-CPA
calculation are stored on a convenient energy mesh, the extra computational
efforts required by the CPA+LF calculation are negligible.

In this paper we discuss results for fcc CuPd and for bcc and fcc CuZn
random alloys at several concentrations. In all the cases we have used the
Local Density approximation (LDA) for the exchange-correlation
potential~\cite{HK&KS}, the ASA approximation for the site potentials and
the angular momentum expansions have been truncated at $l_{MAX}=3$. We use
a fully relativistic treatment for core electrons and a scalar
relativistic approximation for valence electrons. For all the alloy systems
considered in this paper, the lattice parameters have been kept fixed on
varying the concentration. In particular, we set $a=5.5$ a.u. and $a=6.9$ a.u.
for bcc and fcc CuZn, and $a=7.1$ a.u. for fcc CuPd. With this choice, the atomic volumes in
fcc and bcc CuZn alloys differ only about 1.3 per cent. 

\begin{figure}
\centerline{\epsfig{figure=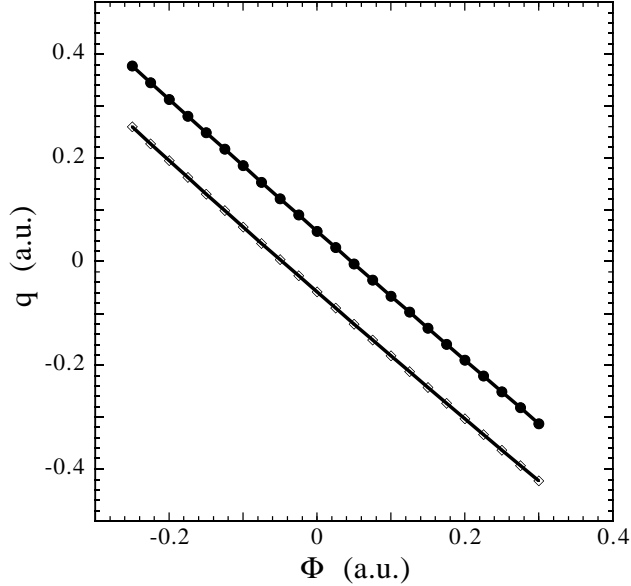,width=8.5cm}}
\vspace{0.5cm}
\caption{Site charge excesses $q_\alpha$ ($\alpha=$ Cu,Zn) vs. the external
field, $\Phi$, from CPA+LF calculations for
Cu$_{0.50}$Zn$_{0.50}$ bcc random alloys. Circles and diamonds,
respectively, indicate Cu and Zn impurities.}
\label{qvsphi}
\end{figure}

As we shall discuss in the next subsection, the charge relaxation occurring at the
impurity site in presence of the external field
phenomena are quite complex. Nevertheless, the CPA+LF model gives a simple
linear relation between the potential $\Phi$ and the
corresponding site charges. In Fig.~\ref{qvsphi} we report $q_\alpha$ ($\alpha=$Cu, Zn) vs. $\Phi$ for
a Cu$_{0.50}$Zn$_{0.50}$ bcc random alloy. As it is evident, the data can
be fitted very well by two straight lines, one for each atomic species (
with correlations that differ from one by less than one part over a
million). Interestingly, the slopes of the two lines are different by a
relatively small but statistically relevant amount, slightly less than 2
 per cent.

\begin{figure}
\centerline{\epsfig{figure=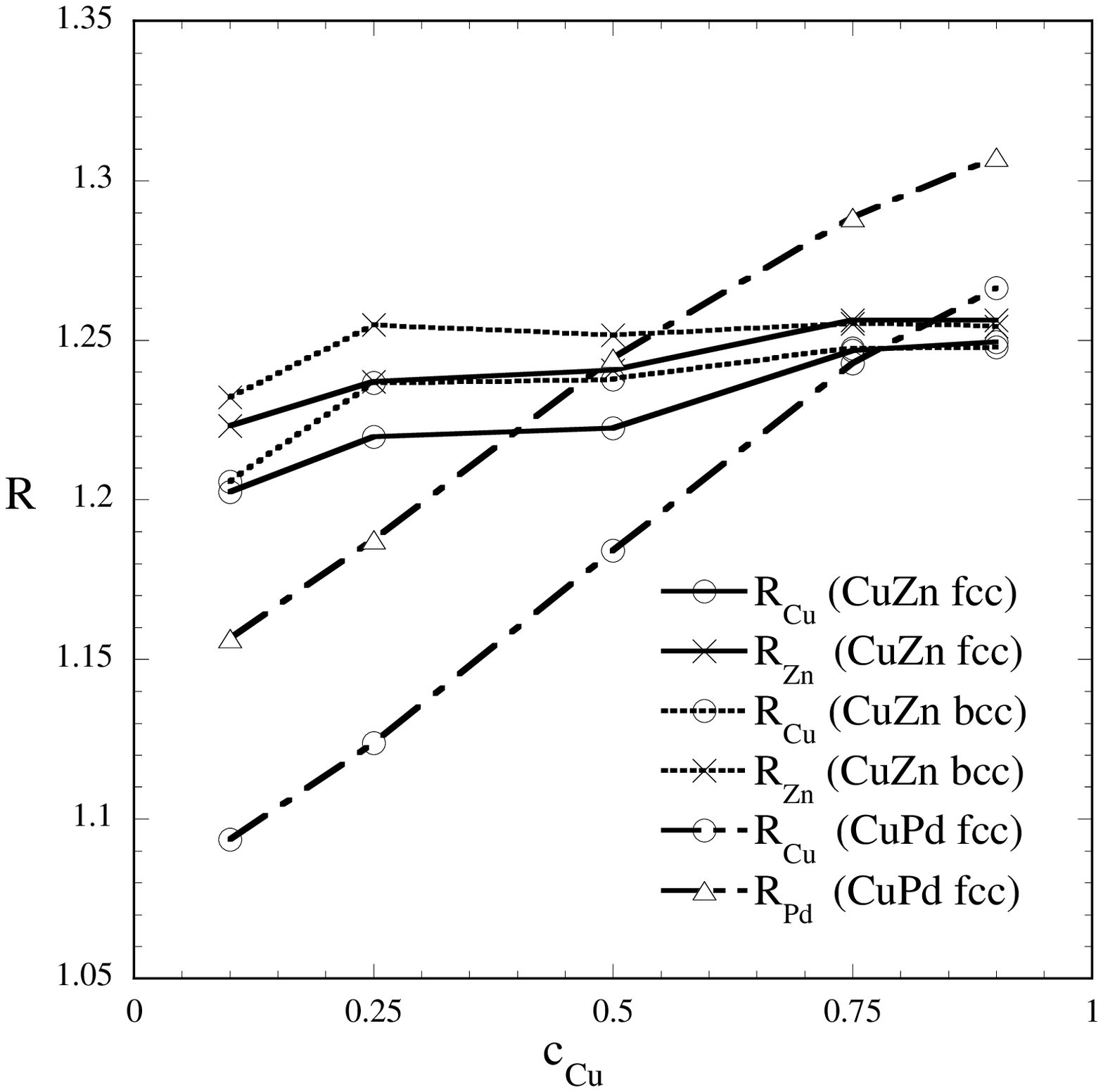,width=7.35cm}}
\centerline{\epsfig{figure=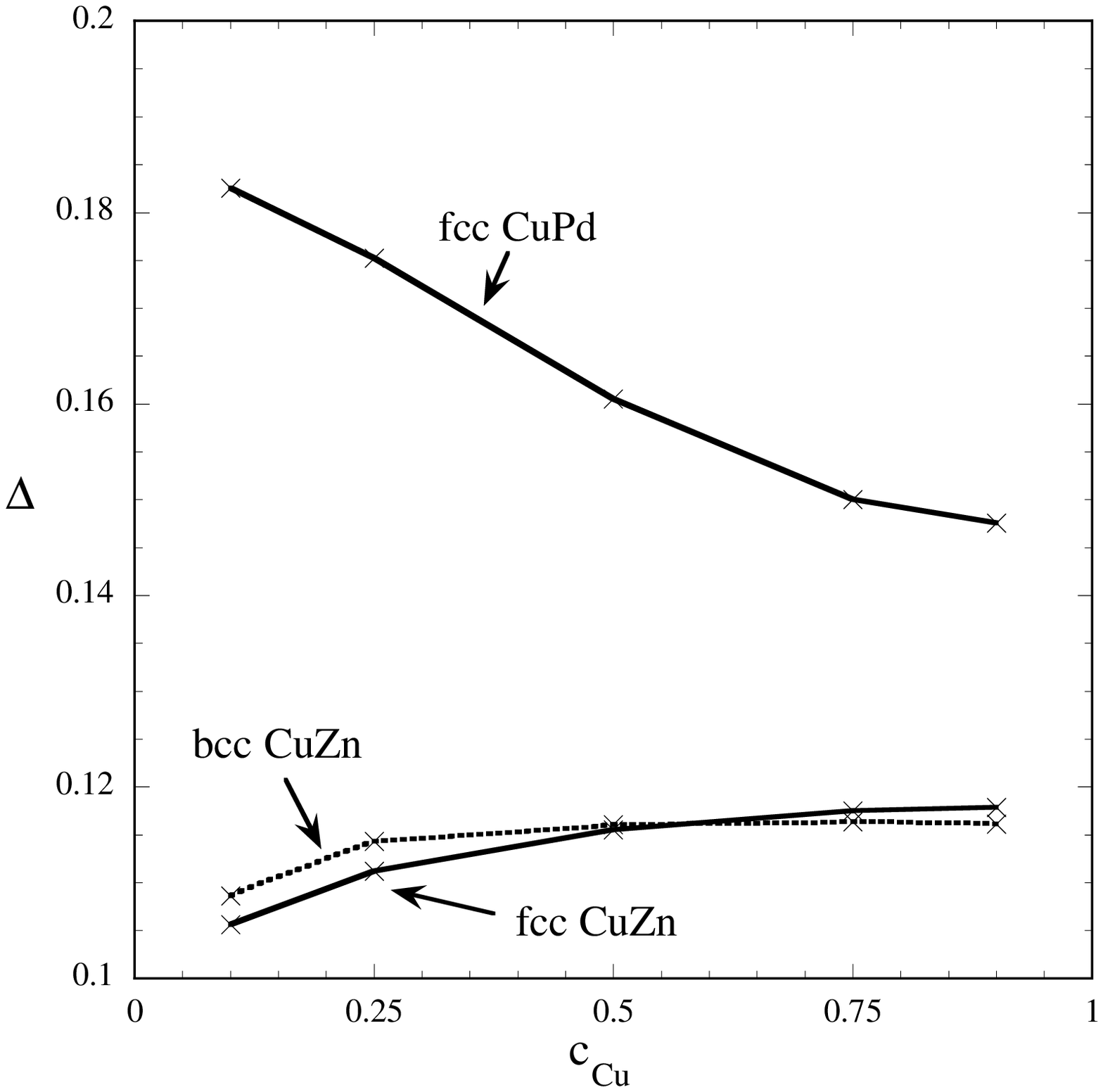,width=7.35cm}}
\vspace{0.5cm}
\caption{Fit coefficients of the linear law $q$ vs. $\Phi$ from CPA+LF
calculations for fcc CuPd and fcc and bcc CuZn alloys plotted vs. the Cu
concent. Upper frame: response coefficients $R_\alpha$, ($\alpha$ refers
to the alloying species); lower frame: 'electronegativity difference',
$\Delta$. The various alloy systems are indicated by labels.}
\label{R_delta_vs_c}
\end{figure}

We notice that in Fig.~\ref{qvsphi} we have considered also $\Phi$ values
considerably larger that those observed in LSMS calculation or
likely to occur in real systems; so according to our data the
linear relations seem to be valid in a quite broad field range.
We have fitted the $q_\alpha$ vs. $\Phi$ curves at
each molar fraction for fcc Cu$_c$Pd$_{1-c}$, fcc Cu$_c$Zn$_{1-c}$ and bcc
Cu$_c$Pd$_{1-c}$ random alloys, at a number of alloy concentrations, using
the linear relationships
\begin{equation}
\label{qfit}
q_\alpha(\Phi)=q_\alpha^0 - R_\alpha \; \Phi
\end{equation}
However, at $\Phi=0$, our CPA+LF model satisfies the CPA 'electronegativity'
condition, Eq.~(\ref{elecn}), and we have:
\begin{equation}
c_A q_A^0 +c_B q_B^0=0
\end{equation}
Henceforth, $q_A^0$ and $q_B^0$ are not independent quantities and we have
chosen as the parameters of our fit only the three quantities $R_A$, $R_B$
and
\begin{equation}
\label{delta}
\Delta = q_A^0/c_B = - q_B^0/c_A = q_A^0  - q_B^0
\end{equation}
The results of these fits are reported in Table 1.
 The trends found for the fitting parameters vs. the alloy
molar fractions are shown in Fig.~\ref{R_delta_vs_c}. The dependence on
the concentration is appreciable for all the fitting
parameters, as expected on the basis of the arguments in section 2. 
Remarkably, the dependences on the alloy system and on the concentration
appear at least as much important as that on the atomic species. Thus, for
instance, for a given alloy system and concentration, there are relatively
small differences between the values of $R$ corresponding to sites occupied
by different atoms. On the other hand, we find much larger variations for $R_{Cu}$
throughout the alloy systems considered. It is
interesting to observe that the trends for the slopes, $R_{Cu}$ and
$R_{Zn}$, and for $\Delta$ are very similar in {\it both} fcc and bcc
CuZn alloys. We notice also that $\Delta$, a measure of the electronegativity
difference between the alloying species, exhibits, at least for CuPd
alloys, non negligible variations vs. the concentration. In the model
of Ref.~\cite{Magri}, the same quantity is assumed independent
on the concentration.
As we see from Table 1, the values for $\Delta$
from our theory are systematically smaller than those from LSMS.
This fact has not to do with the presence of external fields
and it is a feature of the standard CPA theory already discussed in the literature~\cite{BG}. This
notwithstanding, the CPA is able to catch the qualitative
trends of $\Delta$ vs. the concentration for all the systems considered.\\
Although the CPA+LF model gives for $q$ vs. $\Phi$ the same
linear functional form as that obtained for $q$ vs. $V$ from LSMS
calculations, the differences between the two different sets
of calculations forbid, at this stage, a direct comparison of the
fit coefficients. In fact, as we have already stressed, our CPA+LF model
does not account for charge relaxations outside the impurity site volume.
By its construction, the CPA medium C is able to screen the impurity
charge at $\Phi=0$, i.e. $q_\alpha^0$. We can think that this amount of
charge is screened by the infinite volume of C. The introduction of the
local field at the impurity site causes a local excess of charge,
$q_\alpha(\Phi) - q_\alpha^0$,  with respect to the standard CPA. In order
to have global electroneutrality in the CPA+LF theory, it is necessary to
introduce, somewhere outside the impurity site, an opposite amount
of charge, $q_\alpha^0 - q_\alpha(\Phi)$. Here it will be accomplished
 using the arguments of the screened impurity model (SIM-CPA)
model by Abrikosov et al.~\cite{Abrikosov}. We suppose that the excess (with respect to the standard CPA) charge
at the impurity site, $q_\alpha(\Phi)- q_\alpha^0$, is {\it completely
screened} at some distance, $\rho$, of the order the nearest
neighbours distance, $r_1$. Accordingly, in the mean, each of the $n$ nearest
neighbours of the impurity cell has a net
charge excess $(q_\alpha^0-q_\alpha(\Phi))/n$. This, in turn, induces an extra field $\Phi_1 = n (2/\rho)(q_\alpha^0-
q_\alpha(\Phi))/n=2(q_\alpha^0-q_\alpha(\Phi))/\rho$ on the
impurity site.
\begin{table}
\caption{\normalsize{Fit parameters for the $q$ vs. $\Phi$ relationships from CPA+LF
calculations in fcc Cu$_c$Pd$_{1-c}$, bcc Cu$_c$Zn$_{1-c}$ and fcc Cu$_c$Zn$_{1-c}$
random alloys ~\protect\cite{BMZ} . The 'electronegativity
difference', $\Delta$, and the response coefficients, $R_\alpha$,
 are defined in Eqs.~(\ref{qfit}) and ~(\ref{delta}), RMS is the root
mean square deviation. The 'renormalized' response coefficients,
$\tilde{R}_\alpha$ are defined in Eq.~(\ref{qfit_mod}). On the right we report
$\Delta$ and $R_\alpha$ from the LSMS calculations of 
Refs.~\protect\cite{FWS2,FWS1}.}}
\vspace{0.5cm}
\begin{tabular}{cccccccc|ccc}
Alloys & $c$  & $\Delta$ & R$_{Cu}$ & R$_{Pd(Zn)}$ & RMS $\times 10^4$ &
$\tilde{R}_{Cu}$ & $\tilde{R}_{Pd(Zn)}$ &
$\Delta$ & $\tilde{R}_{Cu}$ &  $\tilde{R}_{Pd(Zn)}$  \\
\hline
fcc Cu$_c$Pd$_{1-c}$ & 0.10 & 0.183 & 1.093 & 1.156 & 1.8 & 0.762 & 0.792 & 0.238 & 0.833 & 0.843    \\
& 0.25 & 0.175 & 1.124 & 1.187 & 2.1 & 0.776 & 0.806 & 0.229 & 0.838 & 0.851    \\
& 0.50 & 0.160 & 1.184 & 1.244 & 1.9 & 0.805 & 0.832 & 0.219 & 0.843 & 0.851    \\
& 0.75 & 0.150 & 1.243 & 1.288 & 2.4 & 0.831 & 0.851 & 0.212 & 0.838 & 0.853    \\
& 0.90 & 0.148 & 1.267 & 1.307 & 4.4 & 0.842 & 0.860 & 0.211 & 0.836 & 0.853    \\
\hline
bcc Cu$_c$Zn$_{1-c}$ & 0.10 & 0.109 & 1.206 & 1.232 & 10  & 0.800 & 0.812 & 0.155 & 0.536 & 0.581  \\
& 0.25 & 0.114 & 1.237 & 1.255 & 10  & 0.814 & 0.822 & 0.159 & 0.526 & 0.554  \\
& 0.50 & 0.116 & 1.237 & 1.251 & 6.9 & 0.814 & 0.820 & 0.156 & 0.545 & 0.549  \\
& 0.75 & 0.116 & 1.247 & 1.255 & 5.0 & 0.819 & 0.822 & 0.155 & 0.567 & 0.564  \\
& 0.90 & 0.116 & 1.248 & 1.254 & 3.2 & 0.819 & 0.822 & 0.158 & 0.582 & 0.577  \\
\hline
fcc Cu$_c$Zn$_{1-c}$ & 0.10 & 0.106 & 1.202 & 1.223 & 8.2 & 0.805 & 0.815 & 0.145 & 0.575 & 0.628  \\
& 0.25 & 0.111 & 1.220 & 1.237 & 8.1 & 0.813 & 0.821 & 0.150 & 0.580 & 0.618  \\
& 0.50 & 0.116 & 1.222 & 1.241 & 5.5 & 0.814 & 0.822 & 0.151 & 0.600 & 0.622  \\
& 0.75 & 0.117 & 1.247 & 1.256 & 5.2 & 0.825 & 0.829 & 0.150 & 0.615 & 0.632  \\
& 0.90 & 0.118 & 1.249 & 1.256 & 3.3 & 0.826 & 0.829 & 0.152 & 0.616 & 0.630  \\
\end{tabular}
\label{tabI}

\end{table}
  The total field at the impurity site is then
the sum of the external field $\Phi$ and of the above extra term,
in formulae,
\begin{equation}
\label{screen}
V_\alpha =\Phi + 2 (q_\alpha^0-q_\alpha(\Phi))/\rho
\end{equation}
Then, by solving for $\Phi$ the last equation and substituting in
Eq.~(\ref{qfit}), we find
\begin{equation}
\label{qfit_mod}
q_\alpha(\Phi)=q_\alpha^0 -
\frac{R_\alpha}{1+2R_\alpha/\rho} \; V_\alpha = q_\alpha^0 -
\tilde{R}_\alpha V_\alpha
\end{equation}
The coefficients $\tilde{R}_\alpha$ can be compared directly with
the slopes of the $qV$ relations from LSMS calculations.
However, the comparison, reported in Table 1, requires a caveat:
we have assumed $\rho=r_1$, i.e. a complete screening at the distance of
the nearest neighbours.

 Actually, the screening lengths in metals are of the order
of this distance~\cite{Pines}, but our estimate is too
crude to expect for a very good quantitative agreement with LSMS calculations
in which the charge relaxation is allowed at all the length scales.
However, the agreement found is quite satisfactory, within 10 per cent,
for CuPd alloys, while larger discrepancies are found for CuZn.
Again, the trends for $\tilde{R}_\alpha$ vs. the
concentration are qualitatively reproduced.

\subsection{CPA+LF results for CuZn and CuPd alloys: the charge relaxation}
We have already said, in spite of the $qV$ linear laws, the relaxation
phenomena occurring in presence of an external field are complicated. The CPA+LF model allows for the determination of the response
to an external potential field by the electrons {\it inside} the atomic sphere A.

\begin{figure}
\vspace{1.0cm}
\centerline{\epsfig{figure=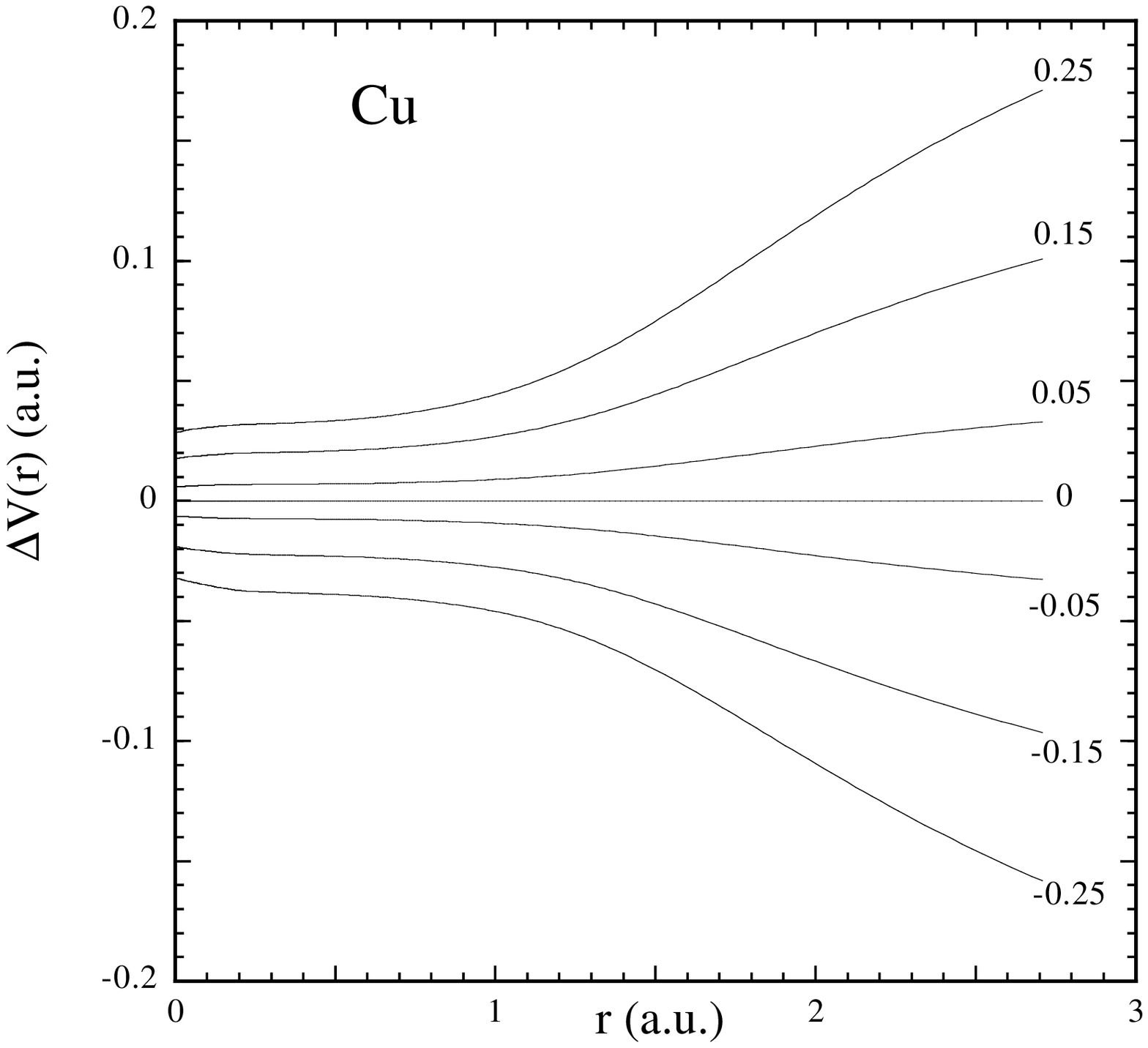,width=6.45cm}}
\vspace{0.5cm}
\centerline{\epsfig{figure=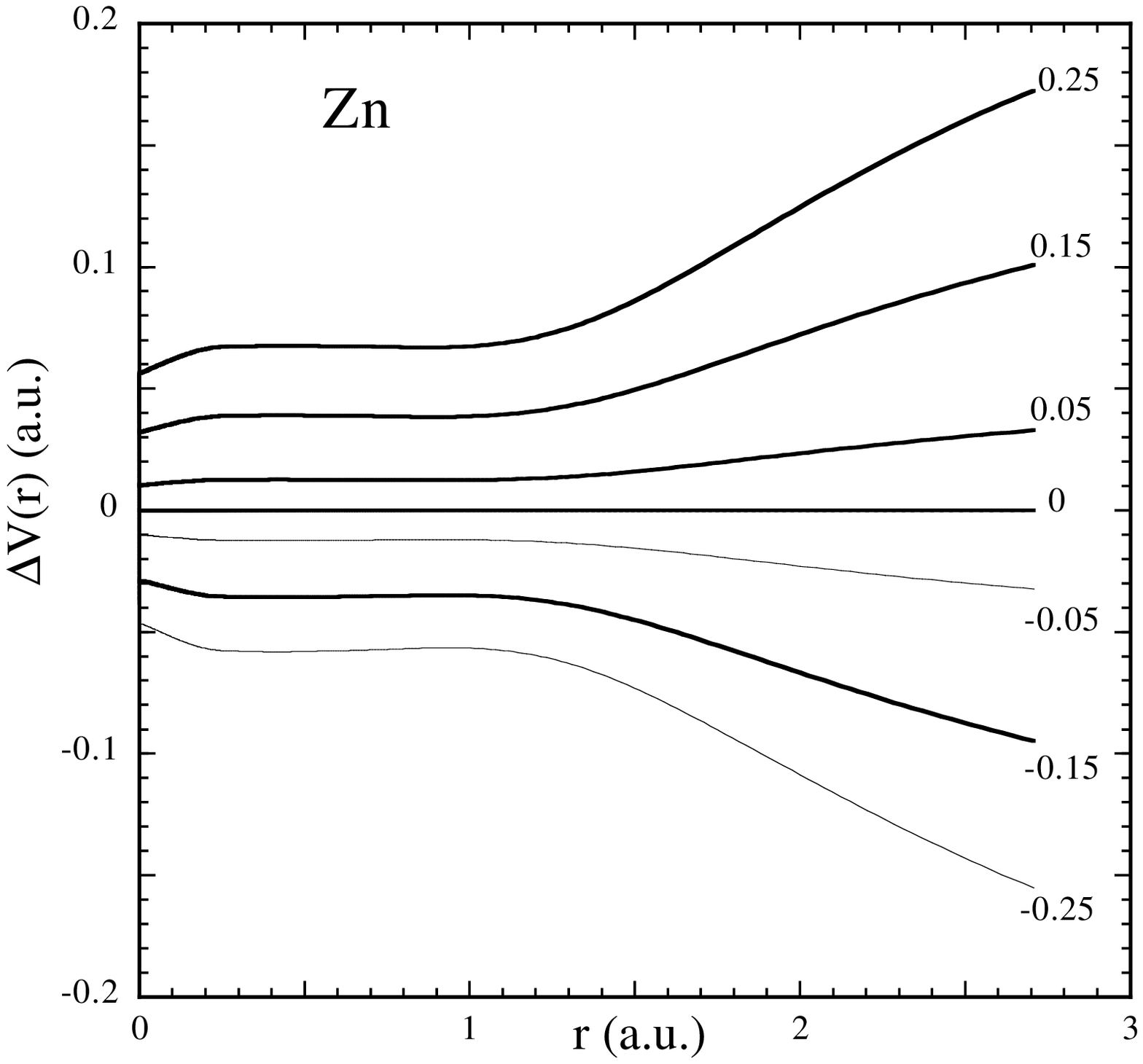,width=6.31cm}}
\vspace{0.5cm}
\caption{Calculated total field $\Delta V^\Phi_\alpha(r)$, $\alpha=$Cu, Zn
(see Eq.~\ref{deltav}) in Cu$_{0.50}$Zn$_{0.50}$ bcc random alloys.
The labels indicate the values of the external field, $\Phi$. At the
Wigner-Seitz radius, $r_{WS} \approx 2.71$ a.u., the total field results to
be about one half of the external field, while the electronic screening is
almost complete at $r=0$.}
\label{vvsphi}
\end{figure}

\begin{figure}
\vspace{1.0cm}
\centerline{\epsfig{figure=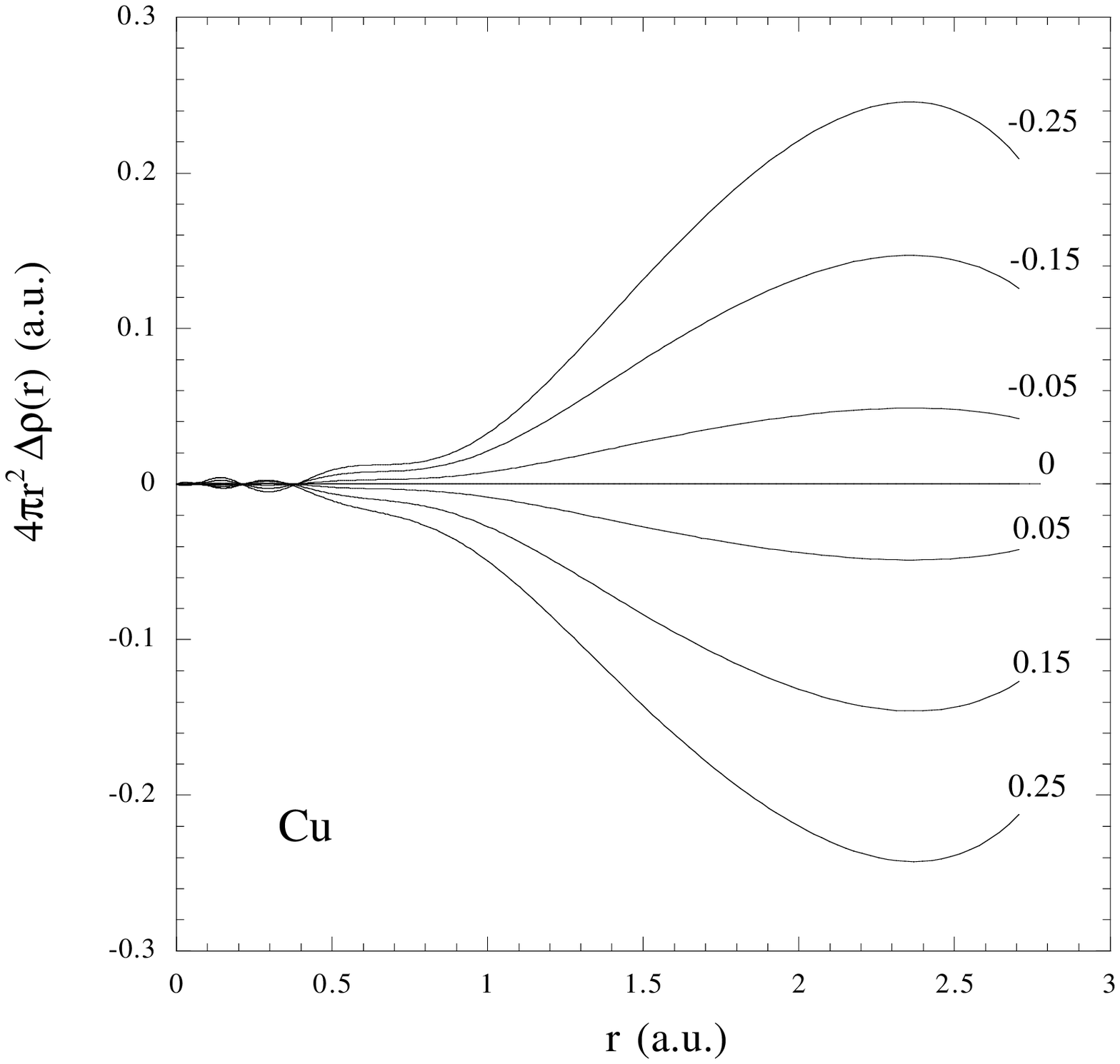,width=6.45cm}}
\vspace{0.25cm}
\centerline{\epsfig{figure=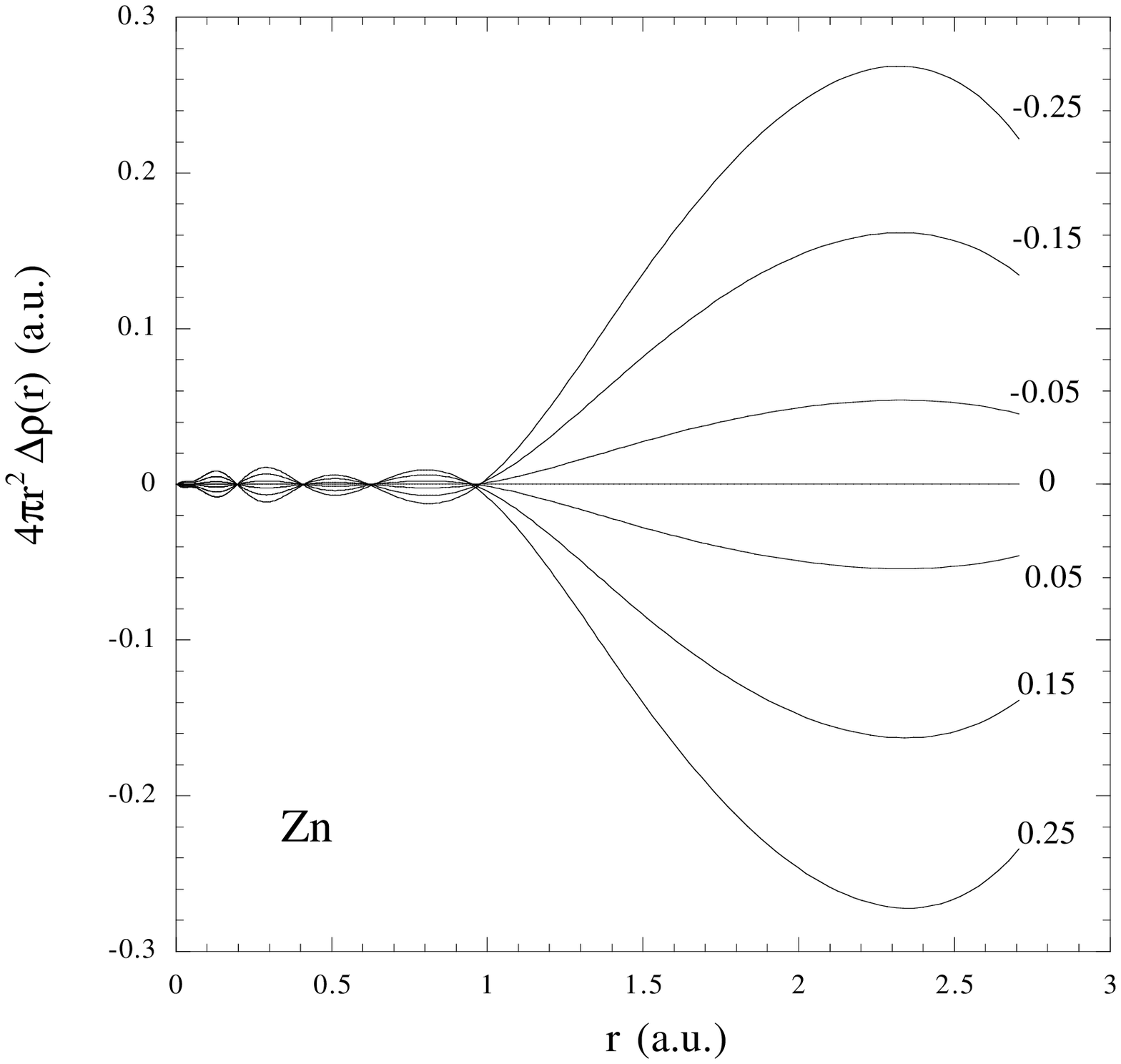,width=6.45cm}}
\vspace{0.5cm}
\caption{Calculated excess charge density, $4 \pi r^2 \Delta \rho^\Phi_
\alpha(r)$ ($\alpha=$Cu, Zn) (see Eq. ~\ref{deltarho}) in Cu$_{0.50}$Zn$_{0.50}$
bcc random alloys. The labels indicate the values of the external field, $\Phi$.}
\label{rhovsphi}
\end{figure}

\twocolumn[\hsize\textwidth\columnwidth\hsize\csname @twocolumnfalse\endcsname
More specifically, the difference
\begin{equation}
\Delta V^\Phi_A(\vec{r})=V^\Phi_A(\vec{r})+\Phi-V^{\Phi=0}_A(\vec{r})
\label{deltav}
\end{equation}
can be
interpreted as the sum of the external field, $\Phi$ and the internal
screening field inside the atomic sphere. Some typical trends for this
quantity are shown in Fig.~\ref{vvsphi}. There we report $\Delta
V^\Phi_\alpha(\vec{r})$, $(\alpha=Cu,Zn)$, for an bcc Cu$_{0.50}$Zn$_{0.50}$
random alloy, that we have selected as a typical case. At the Wigner-Seitz radius,
$r_{WS} \approx 2.71$ a.u., the internal field is able to
screen about one half of the external field, both for Cu and Zn impurities,
while the screening is almost complete for about $r<1$ a.u..
Apparently, the effect of the screening is far from being just a constant
shift of the local chemical potential: if that was the case, in Fig.~\ref{vvsphi}
we would have just equally spaced horizontal lines. What we observe
is much more complicated. For instance we observe that the screening for
small $r$ is greater in Cu than in Zn.
The complex nature of the screening phenomena is further confirmed by a look
at the electronic densities. In Fig.~\ref{rhovsphi} we plot the excess
charge density induced by the external field
\begin{equation}
\Delta \rho^\Phi_A(\vec{r})=\rho^\Phi_A(\vec{r})-\rho^{\Phi=0}_A(\vec{r})
\label{deltarho}
\end{equation}
 both for Cu and Zn sites, again for random bcc Cu$_{0.50}$Zn$_{0.50}$.]


\begin{figure}
\vspace{0.70cm}
\centerline{\epsfig{figure=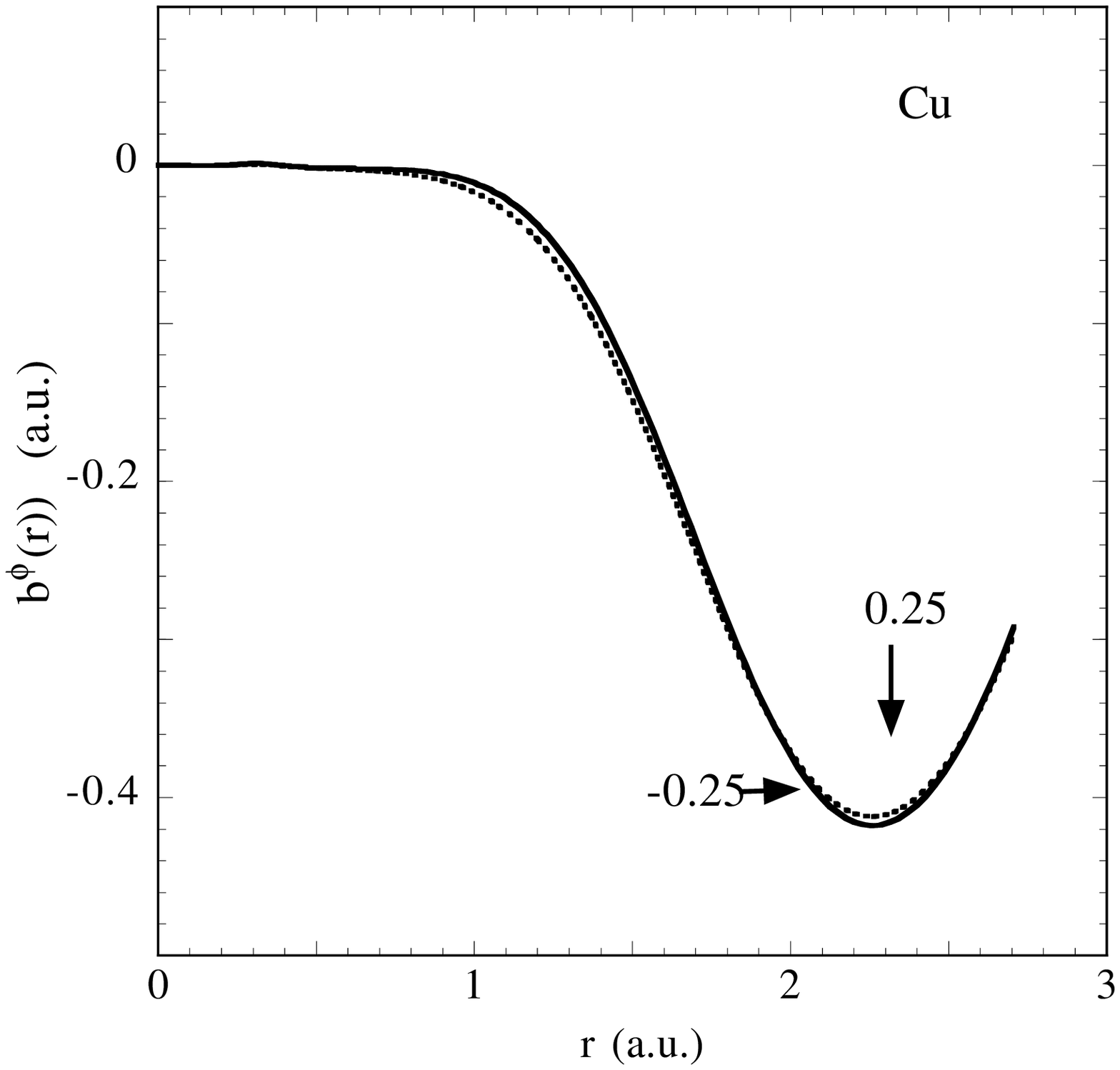,width=6.51cm}}
\centerline{\epsfig{figure=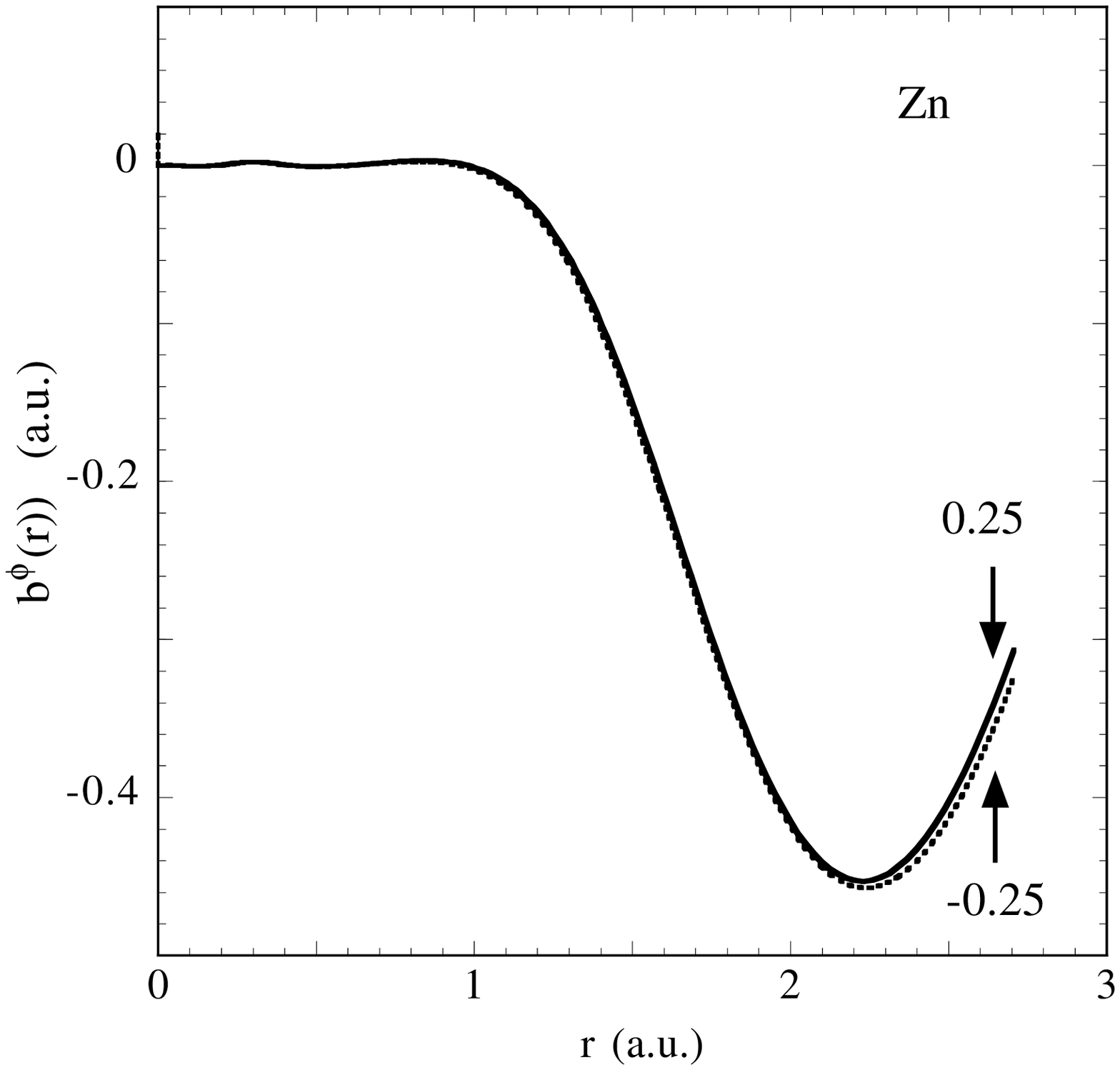,width=6.51cm}}
\vspace{0.5cm}
\caption{ The 'logarithmic derivative' with respect to the external
field (Eq.~\ref{logder}),
$b^\Phi_\alpha(r)$ ($\alpha$=Cu,Zn) in Cu$_{0.50}$Zn$_{0.50}$ bcc random alloys.
The continuous and the dotted lines refer, respectively,
to $\Phi=-0.25$ and $\Phi=-0.25$, the lowest and the highest $\Phi$
values considered.}
\label{rhologvsphi}
\end{figure}

\begin{figure}
\vspace{0.70cm}
\centerline{\epsfig{figure=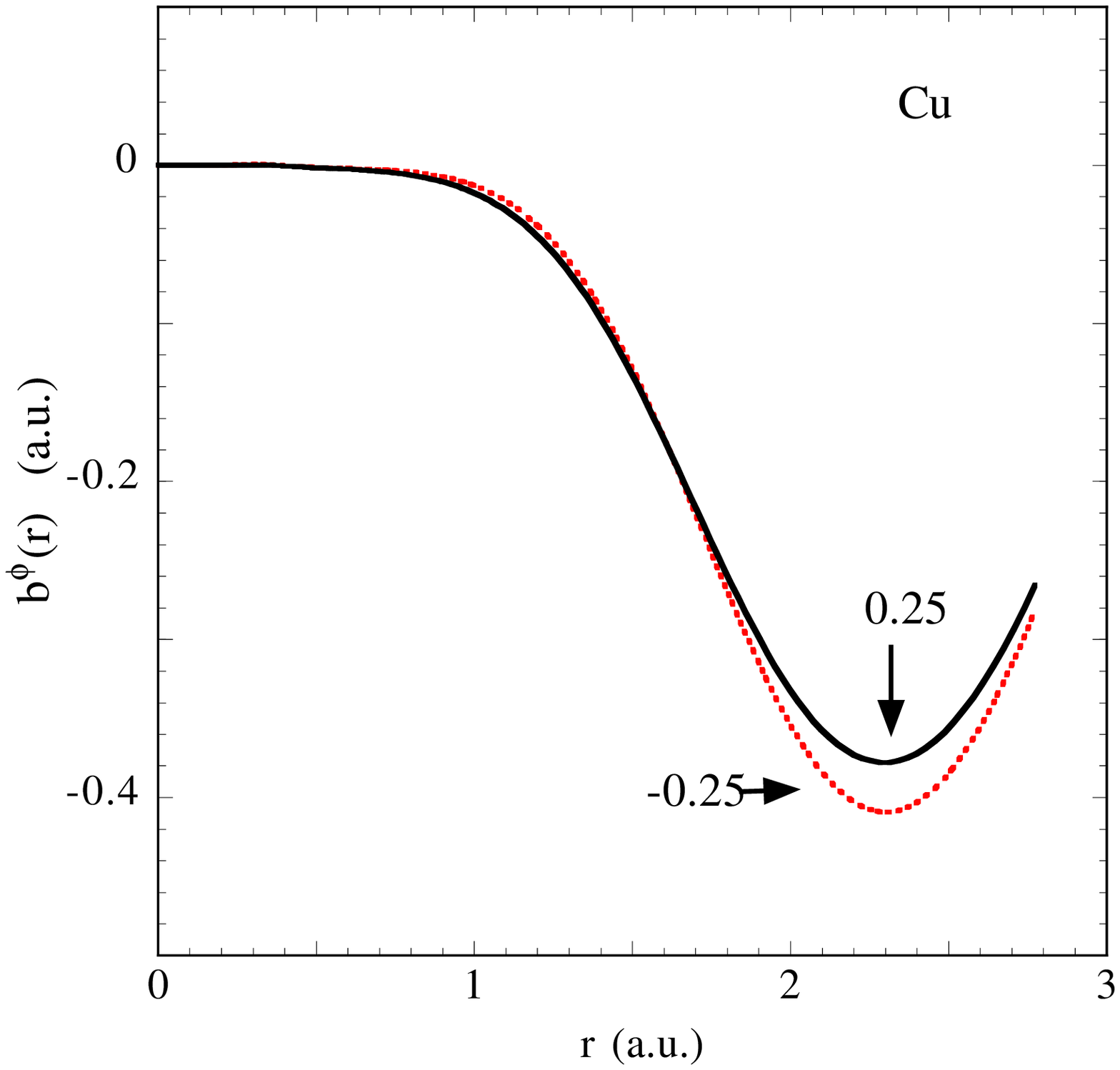,width=6.51cm}}
\centerline{\epsfig{figure=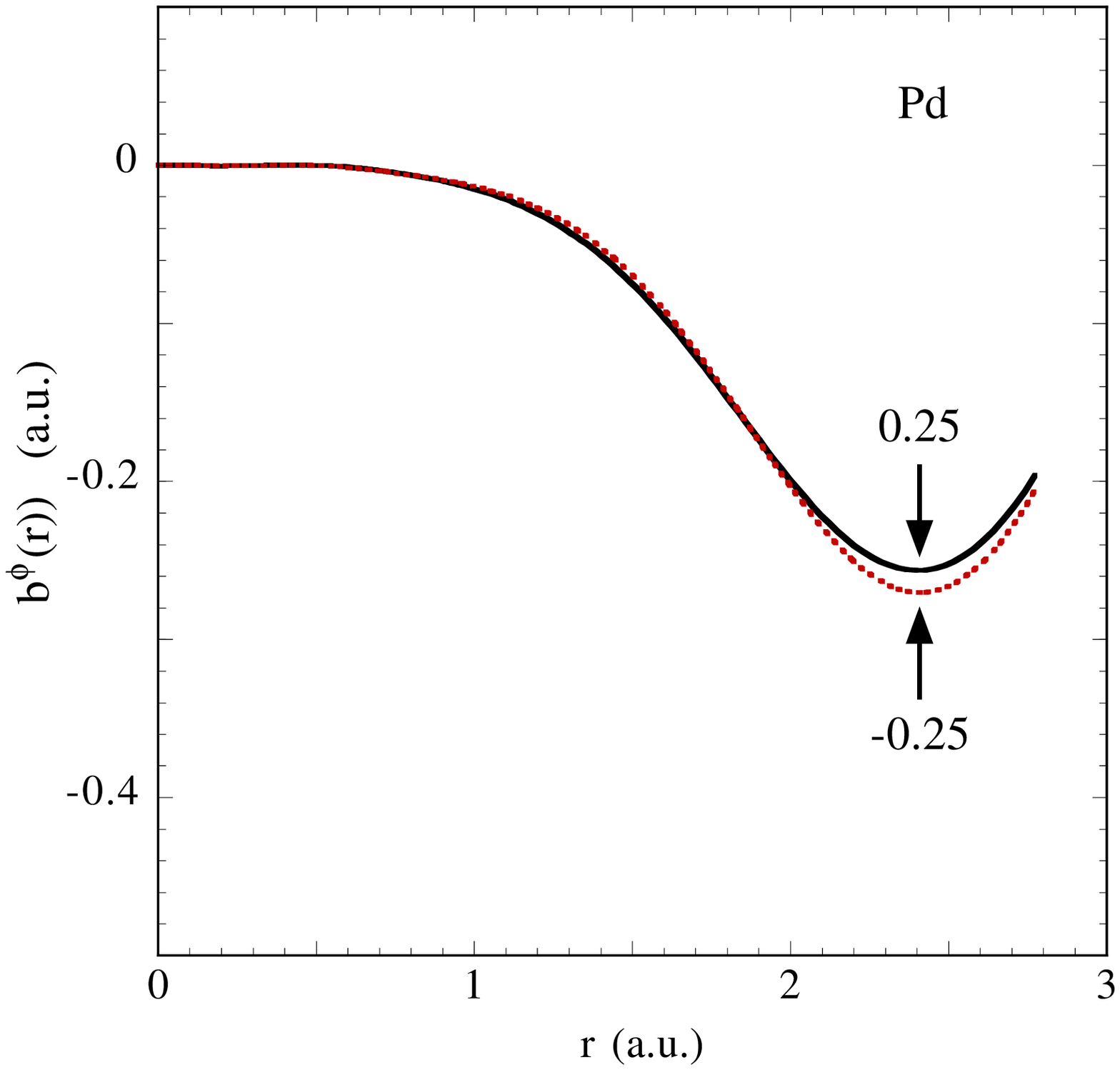,width=6.51cm}}
\vspace{0.5cm}
\caption{ The 'logarithmic derivative' with respect to the external
field (Eq.~\ref{logder}),
$b^\Phi_\alpha(r)$ ($\alpha$=Cu,Pd) in Cu$_{0.50}$Pd$_{0.50}$ fcc random alloys.
The continuous and the dotted lines refer, respectively,
to $\Phi=-0.25$ and $\Phi=0.25$.}
\label{rhologvsphi2}
\end{figure}

\twocolumn[\hsize\textwidth\columnwidth\hsize\csname @twocolumnfalse\endcsname
The largest effects come from the large $r$ region, where the electron density
decreases on increasing $\Phi$ (everywhere in this paper the
expressions "electronic density" or "charge density" are used indifferently
with the meaning of "electron number density", i.e. the charge factor,
$-e$, is {\it not} included). In the innermost part of the atomic spheres,
the variations of the charge density
sometimes may have opposite sign with respect to that observed close to
the cell boundary. We have considered also the quantity,
\begin{equation}
\label{logder}
b_\alpha^\Phi(r)=\frac{\rho_\alpha^\Phi(r)-\rho_\alpha^{\Phi=0}(r)}
{\Phi \;\rho_\alpha^{\Phi=0}(r)} \approx \frac{\partial}{\partial \Phi}
log \rho_\alpha^\Phi(r)
\end{equation}
that, in the limit $\Phi \rightarrow 0$ reduces to the logarithmic
derivative of $\rho_\alpha^\Phi(r)$ and that, on the basis of a formal
scattering theory analysis~\cite{unpub} is expected to have a weak dependence
on $\Phi$. As we can see from Fig.~\ref{rhologvsphi}, where we plot
$b_\alpha^\Phi(r)$ for a bcc Cu$_{0.50}$Zn$_{0.50}$ random alloy
the residual
dependence on $\Phi$ is about a few per cent in a relatively small $r$
interval not far from $r_{WS}$ and less then 1 per cent in
most of the atomic sphere. Moreover this feature appears to be more or less
pronounced depending on the system considered, for this reason we plot in
Fig.~\ref{rhologvsphi2} $b_\alpha^\Phi(r)$ for a fcc Cu$_{0.50}$Pd$_{0.50}$
random alloy. Although the information contained in
Figs.~\ref{rhologvsphi} and Fig.~\ref{rhologvsphi2} can be valuable for the purpose
of improving the
initial guesses for the charge densities, however the dependence of $b_\alpha^\Phi(r)$
on {\it r} appears still quite complicated.
]


\onecolumn

\section{CONCLUSIONS}
The most important result of this work is the reproduction of the linear
laws between local charge
excesses and local electrostatic fields, in good
quantitative agreement with order N electronic structure calculations~\cite
{FWS2,FWS1}.
This is very remarkable if one  considers the {\it single site} nature of our
CPA+LF model, that, hence, requires really modest computational efforts.
The only non first-principles input of our theory has been the inclusion
of a screening length that we have fixed to the nearest neighbours 
distance. Work is in
progress to build a new, completely {\it  ab initio}, version.
The simple mathematical structure of our model has allowed the investigation of
 a range of fields much
broader than that accessible by order N calculations. On this basis, we
can conclude
that the above linear relations have little to do with the {\it size} of the
 external field.
On the other hand, our study shows that, in the same range of fields,
non linear trends
are clearly observable for other site properties, including the charge
density $\rho(r)$
(see, e.g. Figs.~\ref{rhovsphi}, ~\ref{rhologvsphi}, 
~\ref{rhologvsphi2}). \\
As we have already noticed, the CPA+LF theory fixes the reference medium,
 the CPA alloy,
or, in a more mathematical language, the system Green's function that depends only on the
mean molar fractions.  Thus, for a given concentration, any site physical
observable depends only on
the CPA projectors and the site wavefunctions (see Eqs.~(\ref{Green}) and (\ref{tau})),
which, in turn, are completely determined by the nuclear charge on the impurity site and the
coupling potential entering in the corresponding Schroedinger-Kohn-Sham
equation.
Thus, in the CPA+LF theory, {\it any} site property is a {\it unique} function of
the chemical
species and of $\Phi$. A question arises: could this uniqueness be maintained in the more
realistic multiple scattering theory treatment? We argue that, also in
this case, there is
a well defined system Green's function and, in principle
 'site projectors' $D_i$, could be
defined relating the site diagonal part at the site $i$ to
the system Green's function.
The excellent performance of the CPA theory about the spectral properties of
many alloy systems~\cite{Abrikosov_cpa}, the present results and those of Ref.~\cite{Ujfalussy}
suggest that these
 generalized projectors should be very close to their CPA counterparts,
 $D_\alpha$, but, in principle,
they should also be affected by the chemical environment of the first
few neighbours shells of
 $i$-th site. These effects, if they are important, could be studied,
for instance by including
local fields in the charge-correlated CPA scheme by Johnson and Pinski~\cite{more_scr}.
Of course, all the above does not solve the problem of a formal derivation
of the $qV$ laws within the density functional scheme, it simply offers a not too
difficult
mathematical ground in which, we hope, such a derivation could be 
obtained. \\
A further advantage of the CPA+LF model is that, in conjunction with the
 Charge Excess
Functional theory~\cite{BMZc}, it is able to give a good description of
the charge distribution in excellent agreement with order N calculations.
This, and the flexibility
of the scheme, that does not require the use of specific supercells and
is then able to deal
on the same footing with ordered or disordered configurations, suggest
that it constitutes a
first step towards an {\it ab initio} non perturbative theory of
ordering phenomena in metallic alloys.

\begin{acknowledgements}
We thanks Professor J.S. Faulkner and Dr. Y. Wang for having made
available in digital form the data of
Refs.~\cite{FWS2,FWS1}. We acknowledge also discussions with Professor E.S.
Giuliano.
\end{acknowledgements}


\begin{thebibliography}{99}
\bibitem{note}
As noticed in Ref.~\cite{gonissb}, it exists a special value of $r$ at which the charge transfers are
zero for all the alloying species. This is not in contrast with the existence of linear
laws: for the above value of $r$, the range $q_{min} \leq q_i \leq q_{max}$ collapses in
a single point~\cite{Ruban1}. Also in this case $qV$ linear relationships hold with finite
values for $a_A$ and $a_B$ and with $k_A$=$k_B$=0 as it can be established by following the
methods illustrated in section 3.
\bibitem{note2}
This steplike behaviour of the field $\Phi$ in our model corresponds to 
the assumptions made for the Madelung field in atomic sphere or muffin-tin approximation calculations 
for periodic systems.
\end{thebibliography}
\end{document}